\documentclass[prb,twocolumn,amsmath,amssymb,superscriptaddress]{revtex4}
\usepackage{graphicx}
\usepackage{bm}

\begin{document}

\title{Physical Limits of the ballistic and non-ballistic Spin-Field-Effect
Transistor: Spin Dynamics in Remote Doped Structures}
\author{E. Ya. Sherman}
\affiliation{Department of Physics and Institute of Optical Sciences,
University of Toronto, 60 St. George
Street, Toronto, Ontario, Canada M5S 1A7}

\author{Jairo Sinova}
\affiliation{Department of Physics, Texas A\&M University, College Station,
TX 77843-4242, USA}


\begin{abstract}
We investigate the spin dynamics and relaxation in remotely-doped two
dimensional electron systems where the dopants lead to random fluctuations
of the Rashba spin-orbit coupling. Due to the resulting random spin
precession, the spin relaxation time is limited by the strength and spatial
scale of the random contribution to the spin-orbit coupling. We concentrate
on the role of the randomness for two systems where the direction of the
spin-orbit field does not depend on the electron momentum: the spin
field-effect transistor with balanced Rashba and Dresselhaus couplings and
the (011) quantum well. Both of these systems are considered as promising
for the spintronics applications because of the suppression of the
Dyakonov-Perel' mechanism there makes the realization of a spin field effect
transistor in the diffusive regime possible. We demonstrate that the spin
relaxation through the randomness of spin-orbit coupling imposes important
physical limitations on the operational properties of these devices.
\end{abstract}

\maketitle


\section{Introduction}

The simultaneous control of the spin and charge dynamics in low-dimensional
semiconductors by electrical means has been one of the major goals of the
semiconductor spintronics field for over a decade.\cite{Zutic04,Awschalom} A
strong focus within this field has been systems with spin-orbit (SO)
coupling, which connects both the spin and charge degrees of freedom. In
strongly asymmetric heterostructures grown along the [001] direction, the
dominant SO coupling is of the Rashba type due to the inversion asymmetry of
the confining potential, \cite{Rashba84,Rashba60} and has the form ${H}%
_{R}=\alpha_{R}({\sigma}_{x}{p}_{y}-{\sigma}_{y}{p}%
_{x})/\hbar$, acting as a momentum dependent Zeeman field that aligns the
spins of electrons perpendicular to their in-plane momentum. Here, ${%
\sigma}$ are the Pauli matrices, and ${\bm{p}}=({p}_x,{p}_y)$ is
the in-plane momentum. The controlled precession of injected spin polarized
carriers around this effective Zeeman field produced by the Rashba coupling
in momentum space is the basis for the seminal proposal of Datta and Das of
a ballistic spin-field-effect transistor (DD-SFET) \cite{Das}. However, 
a decade long quest to realize such a device has yielded disappointing
results.\cite{Zutic04,Awschalom} In part, this is due to the fact that the
SO coupling, which crucially influences spin transport, also determines the
spin relaxation rate in weakly disordered systems through random precession
due to elastic scattering processes, the so called Dyakonov-Perel' (DP)
mechanism,\cite{Dyakonov72} hence limiting the validity of the
two-dimensional DD-SFET to the ballistic regime.

Schliemann \textit{et al.} \cite{Schliemann02} have recently suggested a way
to design a SFET in (001)-plane based structures that should work on the
diffusive regime, hence without the need to reach this ellusive ballistic
regime, by using the interplay between the Rashba and the Dresselhaus SO
coupling, ${H}_{D}=\alpha _{D}({\sigma}_{y}{p}_{y}-{\sigma}%
_{x}{p}_{x})/\hbar $, which originates from the lack of inversion
symmetry in the unit cell of the host material.\cite{Dyakonov86,Rashba88}
The key aspect of this proposed device is that the momentum dependence of
the direction of the effective Zeeman field is removed if the Rashba and
Dresselhaus coupling strengths are equal with $\alpha _{R}=\pm\alpha _{D}$.
In this case the in-plane SO field is parallel either to the [110] or [1$%
\overline{1}$0] direction and does not depend on the direction of the
electron momentum. It was suggested recently in Refs.[%
\onlinecite{Malshukov00,Kiselev00}] that the freezing of one of the electron
momentum components in one-dimensional channels suppresses the DP mechanism
of spin relaxation and leads to a much longer spin lifetime than in the 2D
quantum wells. This idea, is, however, different from the proposal of Ref.[%
\onlinecite{Schliemann02}], where the DP mechanism becomes much less
efficient through the combined effects of the Rashba and Dresselhaus terms
(see also in Ref.[\onlinecite{Averkiev99}]).

By decoupling the momentum and the spin-quantization axis orientation, the DP
mechanism is greatly reduced, and hence the spin coherence time increased
making such a device apparently realizable in both the diffusive and
ballistic regimes. The condition $\alpha _{D}=\alpha _{R}$ is indeed a
possible one since it has been recently demonstrated that by applying
external bias across the quantum well, one can change the magnitude of $%
\alpha _{R}$ in InGaAs/InAlAs-based \cite{Nitta97,Koga02} and
GaAs/AlAs-based \cite{Knap96,Miller03,Karimov03}, structures and reach
ratios of $\alpha _{D}/\alpha_{R}$ close to one. Another example of the
decoupled SO field and momentum directions is provided by (011) zincblende
quantum wells (QW), where the Dresselhaus SO coupling leads to the effective
field perpendicular to the well plane at arbitrary electron momentum, as
discussed later.

Here we consider an aspect of the practical realization of this device which
will have important limiting consequences in its operating properties. We
show that the randomness of SO coupling arising in doped systems destroys
the regular precession of electron spins and leads to a spin relaxation both
in the SFET device and the (011) quantum wells even in the ballistic regime.
We show that the size of the devices based on the idea of this effective SO
decoupling is limited even in the best possible realizations of the devices
and, therefore, the spins cannot be coherently transferred at arbitrary
distances. In most of the low-dimensional QW-based semiconducting systems,
the SO coupling and finite concentration of electrons are achieved either by
one-side or asymmetric two-side remote doping at distances between the
symmetry plane of the QW and the dopant layer of the order of 10 nm.\cite
{Nitta97,Koga02,Miller03} The spatial non-uniform distribution of dopant
ions leads to fluctuations of the SO coupling, and therefore, the condition of 
$\alpha _{D}=\alpha _{R}$ in a SFET can only be achieved on the average
throughout the sample; the fluctuations of this coupling are analogous to
the classical ''shot'' noise in the transport properties. The effect is, in
some aspects, similar to that investigated by Mel'nikov and Rashba in doped
bulk Si and Ge.\cite{Melnikov72} At the same time, recent theoretical
analysis \cite{Shafir04} shows that in the case of strong SO coupling (large 
$\alpha _{R}$) achieved in highly asymmetric heterosructures the condition $%
\alpha _{D}=\alpha _{R}$ is hardly achievable by applying a bias, and
therefore, we are left with the quantum wells where the moderate Rashba
coupling arises due to the doping or weak structural asymmetry. For a
comparison to possible experimental realizations, we shall consider InGaAs
structures, where the Rashba coupling can be reliably influenced in a wide
range by the applied bias, as has been proven in Refs.\cite{Nitta97,Koga02}.

We organize the rest of the article as follows: In Sec. II we discuss the
properties of the SO coupling considered. In Sec. III, we discuss the spin
relaxation in the proposed field-effect (FET) with balanced Rashba and Dresselhaus SO
coupling terms and in Sec. IV, the one for the (011) quantum well. In Sec. V,
we present conclusions for these results.

\section{Randomness of the Zeeman SO field.}

We consider a two-dimensional (2D) channel and a very thin dopant layer ($%
\delta $-doping) separated by a distance $z_{0}$ with a two-dimensional
concentration $n(\mathbf{r})=\sum_{j}\delta (\mathbf{r}-\mathbf{r}_{j})$ of
dopants at points $\mathbf{r}_{j}$ with charge $e$ and $\mathbf{r}%
=(r_{x},r_{y})$ being the 2D in-plane vector.

Since we are interested in quantum wells, where the Rashba SO coupling
arises due to an asymmetric doping, we assume that the effect of the
electric field of dopants on the SO strength is linear in the $z-$component
of the field ${E}_{z}({\bm \rho })$, that is $\alpha _{R}({\bm \rho }%
)=\alpha _{\mathrm{SO}}e{E}_{z}({\bm\rho })$. Here ${\bm \rho }$ is a point
at the plane of a sufficiently narrow 2D channel and $\alpha _{\mathrm{SO}}$
is a phenomenological system-dependent parameter \cite{Silva97}. This
relationship takes into account the linear in ${E}_{z}$ polarization of the
electron wavefunction as well, as found in Ref. [\onlinecite{Grundler00}]. In
more asymmetric heterostructures or asymmetrically grown quantum wells, the
structural asymmetry will lead to a regular coordinate-independent shift $%
\alpha _{0}$ to the Rashba term, which has no influence on the effects of
the randomness. The $z$-component of the Coulomb field due to the dopant
ions is given by 
\begin{equation}
{E}_{z}({\bm \rho })=\frac{{e}z_{0}}{\epsilon }\sum\limits_{j}\frac{1}{[({%
\bm \rho }-\mathbf{r}_{j})^{2}+z_{0}^{2}]^{3/2}},
\end{equation}
where $\epsilon $ is the dielectric constant, and the summation is performed
over the dopant layer.

Fig 1. presents the pattern of $\alpha _{R}({\bm \rho })$ at the spatial
scale of 200 nm for $z_{0}=20$ nm obtained by a Monte-Carlo produced
''white-noise'' distribution of the dopant ions \cite{Pershin05}. The
fluctuations of $\alpha _{R}({\bm\rho })$ correlated on the spatial scale of
the order of $z_{0}$ become smaller and smoother with the increase of the
distance to the dopant layer, as is shown by one-dimensional cuts
in Fig.2.

The Rashba parameter is a sum of the mean value produced by the asymmetric
field $\langle \alpha \rangle $ and a random contribution with zero mean: 
\begin{equation}
\alpha _{R}({\bm \rho })=\langle \alpha \rangle +\alpha _{\mathrm{r}}({\bm %
\rho }),
\end{equation}
with $\langle \alpha \rangle =\alpha _{0}+2\pi \alpha _{\mathrm{SO}}\left(
e^{2}/\epsilon \right) \overline{n},$ where $\overline{n}$ is the mean value
of the 2D dopant concentration. For the $\delta (z+z_{0})-$doping in quantum
wells with $\alpha _{0}=0$ the relative amplitude of the fluctuations $\sqrt{%
\left\langle \alpha _{\mathrm{r}}^{2}\right\rangle }/\langle\alpha\rangle
=1/\sqrt{8\pi \overline{n}z_{0}^{2}}$. \cite{Efros89,Sherman03a} In contrast
to the regular term, which demonstrates a dependence $\alpha _{R}(V)$ on the
applied  bias $V$ across the well, this random contribution remains
relatively weakly influenced and, therefore, cannot be removed.

\section{Spin relaxation in the non-ballistic SFET condition.}

The randomness in the direction of the spin precession axis can be removed
when at some applied bias $V$, the condition $\alpha _{D}=\pm \alpha _{R}(V)$
is satisfied and $\alpha _{R}(V)$ is spatially uniform. Therefore, the
direction of the linear in the in-plane momentum Zeeman field does not
depend on the momentum directions, and the spin precession becomes more
regular. Here a new integral of motion $\Sigma =\langle \sigma _{x}+\sigma
_{y}\rangle /\sqrt{2}$ appears and supports the regularity of spin dynamics. 
\cite{Schliemann02} We emphasize here that the exact matching is possible
only at the limit of zero electron momentum, where both the Rashba and
Dresselhaus Hamiltonians are linear in $p$. At finite momenta the
Dresselhaus Hamiltonian must be supplemented by the bulk $k^{3}$-like term,
which in  the (001) quantum well has the form: $\alpha _{c}{k}_{y}{k}_{x}(%
{\sigma}_{x}{k}_{y}-{\sigma}_{y}{k}_{x})$, where $\alpha _{c}$
is the Dresselhaus constant for the bulk, and ${\bm k}={\bm p}/\hbar $ is
the wavevector. The role of this term depends considerably on the quantum
well width. This effect, as well as the possible but not yet well known momentum
dependence of the Rashba parameter, can hamper the ability of design the
SFET even if the linear contributions are balanced. 

The total Hamiltonian describing the system written in the Hermitian form:

\begin{widetext}

\begin{eqnarray}
{H}_{\mathrm{SO}} &=&{H}_{D}+{H}_{R},  \\
{H}_{R} &=&{\left\langle \alpha _{R}\right\rangle}
({\sigma}_{x}{k}_{y}-{\sigma}_{y}{k}_{x})+ 
\frac{1}{2}\left[\sigma _{x}\left\{{k}_{y},\left(\alpha_{R}\left({\bm\rho}\right)-\left\langle\alpha _{R}\right\rangle\right)
\right\} -\sigma _{y}\left\{{k}_{x},\left( \alpha _{R}\left( {\bm \rho }%
\right) -\left\langle \alpha _{R}\right\rangle \right) \right\} \right] , \nonumber \\
{H}_{D} &=&\alpha _{D}({\sigma}_{y}{k}_{y}-{%
\sigma}_{x}{k}_{x})+{\alpha _{c}}{k}_{y}k_{x}({%
\sigma}_{x}k_{y}-{\sigma}_{y}k_{x}),  \nonumber
\end{eqnarray}
\end{widetext}
where $\left\{{k}_{i},\alpha _{R}\left({\bm\rho}\right)\right\} $
stands for an anticommutator. The Dresselhaus coupling constant $\alpha
_{D}=\alpha _{c}\langle k_{z}^{2}\rangle $, where $\langle k_{z}^{2}\rangle $
is the expectation value of the $k_{z}^{2}=-{\partial ^{2}}/{\partial z^{2}}$
operator in the ground state, and $\alpha _{c}\approx 25$ eV\AA $^{3}$ in
GaAs and InAs-based structures \cite{Winkler}. In the case of the rigid
walls of the QW with the width $w$, $\langle k_{z}^{2}\rangle =(\pi /w)^{2}$.

Both the cubic terms in the Hamiltonian and the randomness of SO coupling
lead to the spin relaxation in the SFET device. Our main interest in this
paper is the role of the randomness. However, to find the regime where the
effect of the randomness dominates, we will first discuss the contribution
of the $\alpha_{c}{k}_{y}k_{x}({\sigma}_{x}k_{y}-{\sigma}_{y}k_{x})$ 
term for the spin relaxation in the quantum well in the
case $\alpha _{D}=\left\langle \alpha _{R}\right\rangle .$ We will use the
approach developed in Refs. \cite{Dyakonov72,Dyakonov86,Averkiev99}, where
spin relaxation is described by: 
\begin{equation}
\frac{d\sigma _{i}}{dt}=-\sum_{j}\Gamma _{ij}\sigma _{j},
\end{equation}
with $\Gamma _{ij}$ being the anisotropic spin relaxation tensor. For the
relaxation of  $\Sigma $ (expected to be conserved in the $\alpha
_{D}=\left\langle \alpha _{R}\right\rangle $ case) we obtain:

\begin{equation}
\frac{d\Sigma }{dt}=-\Gamma _{D}\Sigma ,
\end{equation}
where the contribution of the $k^{3}-$originated terms yields $\Gamma_{D}=\alpha_{c}^{2}k^{6}\tau
_{k}/8\hbar ^{2},$ with $\tau _{k}$ being the momentum relaxation time. We
will demonstrate that under realistic conditions, this term is much smaller than
the contribution of randomness of the SO coupling, which we consider next.

The randomness of the SO coupling discussed here makes the precession
irregular and, therefore, pushes the range of parameters in which such a
SFET can be realized toward the ballistic regime. We consider a SFET
configuration where electrons are injected through perfect nanoscale quantum
point contacts \cite{Schliemann02} located at points $x=0$, $y=y_{i}$, where 
$i=1,2,\ldots N$ numerates the contacts and the electrons and then move
along the $x$ axis as shown in Fig.1 interacting with the random SO field.
The initial spin state of all electrons is 
\begin{equation}
\phi _{i}(t=0,x=0)=\frac{1}{\sqrt{2}}\left( 
\begin{array}{l}
1 \\ 
e^{i\pi /4}
\end{array}
\right) ,
\end{equation}
with all the spins initially polarized along the ${x}+{y}$
direction. The shape of the injected electron density along the $y$-axis, $%
|\psi (y-y_{i})|^{2}$ formed by the quantum contact is conserved during the
electron propagation such that the corresponding wavefunction has the form: $%
\Psi _{i}(x,y)=\psi (y-y_{i})\exp (ik_{x}x)$. We assume that the $y$-axis
spatial distribution of the wavefunction is much less than $z_{0}$, and,
therefore, the spin of the $i$th electron interacts with the local Rashba
field $\alpha _{R}(x,y_{i})$ and its spatial derivatives. To estimate the
maximum spin-coherence lifetime we assume that the movement is
ballistic, that is the momentum of injected electron is conserved, and,
therefore, only the randomness of $\alpha _{R}(x,y)$ leads to the spin
decoherence, thus, presenting the spin relaxation mechanism in relatively
high-mobility structures with the electron free path much larger than 10 nm.
This restriction implies that the electron kinetic energy is much larger
than the potential fluctuations and ignores quantum mechanical interference
effects which are small in the larger system sizes considered here.
Relaxation of this approximation will further decrease the spin-coherence
lifetime in the diffusive regime.

Under these conditions the random contribution to the SO Hamiltonian
obtained with Eq.(3) is: 
\begin{equation}
{H}_{\mathrm{r}}=-\alpha_{\mathrm{r}}({\bm \rho}){\sigma}_y k_x+ 
\frac{i}{2} \left( {\sigma}_y\frac{\partial\alpha_{\mathrm{r}}({\bm\rho})%
}{\partial x}- {\sigma}_x\frac{\partial\alpha_{\mathrm{r}}({\bm\rho})}{%
\partial y} \right).
\end{equation}

The spin of a moving electron is, therefore, a subject of a randomly
time-dependent SO field, with each electron probing the realization of the
field at the given path or different electrons probing the different
configurations of essentially the same random field. The direction and the
magnitude of the Zeeman field randomly change in time, and therefore, a
dephasing of the spin states occurs. The correlation function of the random
field is characterized by the Fourier component $\langle \alpha _{\mathrm{r}%
}(0)\alpha _{\mathrm{r}}(\tau )\rangle _{\omega }$ extended in the range of
frequencies $\omega $ up to $\Omega _{d}\sim 1/\tau _{d}$ where $\tau
_{d}=z_{0}/v$ is the time the electron takes to travel a distance $z_{0}$
through which the Rashba coupling changes appreciably, and $v$ is the
electron velocity. The frequency $\Omega _{d}$ is much larger than the
spin-flip frequency $\Omega _{\mathrm{SO}}\sim \alpha _{R}k/\hbar $ with $%
\Omega _{\mathrm{SO}}/\Omega _{d}$ being of the order of 0.1. This condition
implies that the random contribution to the Zeeman field is out of resonance
with the spin-flip transitions, and, for this reason, the spin-flip
transition rate is low. Therefore, under these conditions it is necessary to
pass through many domains of the SO coupling for $\Sigma $ to be destroyed.
Analysis similar to that done in Refs.[\onlinecite{Sherman03a,Glazov05}]
shows that in the case $\Omega _{d}\gg \Omega _{\mathrm{SO}}$ the random SO
coupling mechanism leads to the spin relaxation rate of the order of $2\pi
\langle \alpha _{\mathrm{r}}^{2}\rangle z_{0}\mathrm{max}(k^{2},z_{0}^{-2})/{%
\hbar }^{2}v$. In the system considered here, with $k^{2}\gg z_{0}^{-2}$ it
scales with the system parameters as $\langle \alpha _{{R}}^{2}(V=0)\rangle /%
\overline{n}z_{0}k$. This trend is seen in Fig.3 which presents the results
of the Monte Carlo simulation of the spin dynamics. As one can see in Fig.
3, $(\langle\sigma_{x}\rangle+\langle\sigma_{y}\rangle)/\sqrt{2}$ is
gradually destroyed at the electron path $L_{s}$ of the order of 50 $\mu$m, that is about $5\times{10^{3}}$
domains of the SO random coupling. We associate this decay time with the
spin-coherence lifetime in the system. The role of the randomness is
demonstrated by the fact that the spin-coherence lifetime depends very
strongly on the distance between the dopant layer and the conducting
electrons increasing with the distance $z_{0}$. Even when the mean SO
coupling is the same, the increase in the distance between the dopant layer
and the symmetry plane of the quantum well decreases the randomness and, in
turn, the spin relaxation rate, as can be seen from a comparison of the $%
\Sigma (t)$ dependences in Fig.3. The noise in the $\Sigma (t)$%
-dependence corresponds to the simulation with a finite number of electrons
and must be experimentally observable in the SFETs operating in
this mode. For these reasons, the size of the transistor base cannot be very
long, and the spins cannot be controllably delivered at distances much
larger than 100 $\mu $m at best.

An order of magnitude estimate of the ratio of the
relaxation rates due to these two mechanisms, namely, the SO coupling
randomness $\Gamma _{\mathrm{r}}$ and the $k^{3}-$terms contributions 
$\Gamma _{D}$ gives: 
\begin{equation}\label{eq::Gammaratio}
\frac{\Gamma _{\mathrm{r}}}{\Gamma _{D}}\sim 16\pi \frac{\left\langle \alpha
_{\mathrm{r}}^{2}\right\rangle }{\langle \alpha \rangle _{R}^{2}}\left( 
\frac{\pi }{wk}\right) ^{4}\frac{\tau _{d}}{\tau _{k}}.
\end{equation}
The small ratio $\tau _{d}/\tau _{k},$ which strongly depends on the
mobility, is typically of the order of $10^{-2},$ and, therefore, favors the
role of the $k^{3}$ Dresselhaus contribution. However, at $wk<1,$ it is
compensated by the $\pi^{4}$ prefactor arising due to the size quantization
and by the effect that due to the fast in-plane angular dependence of the $k^{3}-$originated
terms, the efficiency of this mechanism is decreased,
leading to a large prefactor $16\pi $ in Eq.\ref{eq::Gammaratio}. The exact value of this
prefactor is model-dependent, being, however, of the order of ten at any
reasonable model. Therefore, even at relatively small fluctuations of the
Rashba SO coupling $\left\langle \alpha _{\mathrm{r}}^{2}\right\rangle
/\langle \alpha \rangle _{R}^{2}\sim 0.1$, they will dominate as the spin
relaxation mechanism. We mention here that despite the spin-orbit coupling
in our model depending linearly on the momentum, in the quasiballistic regime
the spin relaxation path $L_{s}=\hbar k/m\Gamma _{\mathrm{r}},$ where $m$ is
electron effective mass, shows only a weak electron-momentum dependence. The
reason is that the random contribution to the spin precession angle for
passing through one domain, being of the order of $\alpha _{\mathrm{r}%
}k/\hbar \times z_{0}/(\hbar k/m)$, is momentum-independent, and, therefore,
the number of the domains contributing into $L_{s}$, depends weakly on $k.$
At the same time, for the Dresselhaus $k^{3}-$contribution one would expect 
$L_{s}\sim k^{-5}.$

\section{The \textrm{(011)} quantum well}

Another interesting system where the direction of the SO field is
momentum-independent is the (011) zincblende quantum well. Here SO coupling
is described by the Dresselhaus Hamiltonian 
\begin{equation}
{H}_{D}=\alpha _{D}k_{y}{\sigma}_{z}\left[ 1-\left(
k_{y}^{2}-2k_{x}^{2}\right) /\langle k_{z}^{2}\rangle \right] ,
\end{equation}
with the direction of the SO field being always parallel to the $z$-axis
which is perpendicular to the QW plane. Here $\alpha _{D}=\alpha _{c}\langle
k_{z}^{2}\rangle /2.$ The $\left( k_{y}^{2}-2k_{x}^{2}\right) /\langle
k_{z}^{2}\rangle $ term in the square brackets is the bulk-originated Dresselhaus
term, which, contrary to the (001) quantum well, due to the symmetry, does
not lead to the change in the SO field direction with the changes in the
momentum. When the system is the subject of the regular Rashba Hamiltonian 
${H}_{R}=\left\langle\alpha_{R}\right\rangle({\sigma}_{x}{k}_{y}-{\sigma}_{y}{k}_{x})$ 
also, the electron
spin precesses around the axis determined by the direction of the 
${H}_{D}+$ ${H}_{R}$ field with the rate $\Omega =2\sqrt{\alpha
_{D}^{2}k_{y}^{2}+\alpha _{R}^{2}k_{\parallel }^{2}}/\hbar ,$ where $%
k_{\parallel }^{2}=k_{x}^{2}+k_{y}^{2}.$ For the component of the spin
initially polarized along the $z-$axis, one obtains

\begin{eqnarray}
\sigma _{z}(t) &=&\overline{\sigma }_{z}+A_{\sigma }\cos \Omega t, \\
\overline{\sigma }_{z} &=&\frac{\alpha _{D}^{2}k_{y}^{2}}{\alpha
_{D}^{2}k_{y}^{2}+\alpha _{R}^{2}k_{\parallel }^{2}}, \nonumber \\
A_{\sigma } &=&\frac{\alpha _{R}^{2}k_{\parallel }^{2}}{\alpha
_{D}^{2}k_{y}^{2}+\alpha _{R}^{2}k_{\parallel }^{2}}, \nonumber 
\end{eqnarray}
where $A_{\sigma }$  is the precession amplitude. Randomness in ${H}_{R}$
causes fluctuations in the magnitude and direction of the Zeeman SO field
acting on the electron spin, both now changing  randomly in time. Eventually,
this leads to the spin relaxation, which can be also understood as the ${%
\sigma }_{z}=1$ to ${\sigma }_{z}=-1$ spin-flip transitions caused by the
random field.

We consider the case of the asymmetrically doped quantum well for which the mean
value of the Rashba parameter can be influenced by applied bias. A typical
evolution of $\left\langle \overline{\sigma }_{z}(t)\right\rangle $ averaged
over an ensemble of electrons for the parameters of spin-orbit coupling typical
for In$_{0.5}$Ga$_{0.5}$As quantum wells is presented in Fig. 4. We consider
two cases: a partially and the fully compensated Rashba terms. In the former
case the relaxation occurs as both the amplitude of the oscillations and the
mean value averaged over the ensemble of electrons tend to zero. As one can
see in the Fig. 4, the decay time of $\langle \sigma _{z}(t)\rangle $ in
the In$_{0.5}$Ga$_{0.5}$As structures is of the order of 10 ps, similar to
the case of the $\alpha _{D}=\langle \alpha _{R}(V)\rangle $ transistor.

In ''pure'' GaAs structures, where $\alpha _{\mathrm{SO}}$ is an order of
magnitude smaller than in the In$_{0.5}$Ga$_{0.5}$As alloys, one would
expect the spin relaxation time due to the randomness of the doping which
roughly scales as $\alpha _{\mathrm{SO}}^{-2},$ two orders of magnitude
longer, that is of the order of 10$^{3}$ ps. A long electron spin coherence
time $T_{s}$ of this order of magnitude has indeed been observed
experimentally in Refs.[\onlinecite{Ostreich04}] for the (011) symmetrically
doped GaAs QWs of relatively low mobility with $\mu \sim 10^{3}$ cm$^{2}$%
/(Vs). It is instructive to estimate the corresponding distance $L_{s}$ at
which the spins of diffusively propagating electrons can be delivered for
this time in these low-mobility samples. With this mobility, the
momentum relaxation time $\tau \sim 0.1$ ps leads to $L_{s}\sim v(T_{s}\tau
)^{1/2}$, where $v\sim 10^{7}$ cm/s is the speed of optically injected
electrons investigated in Ref.[\onlinecite{Ostreich04}]. Here $(T_{s}\tau
)^{1/2}\sim 10$ ps, and therefore, $L_{s}$ is of the order of a few $\mu 
\mathrm{m}$. Therefore, due to the low mobility these QWs cannot be used
for the simultaneous long-distance charge and spin transfer necessary for
the spintronics applications despite a long spin coherence time there.

\section{Conclusions}

We have investigated the spin relaxation rate arising due to a random SO
coupling in a possible realizations of a SFET with the balanced
Rashba and Dresselhaus spin-orbit coupling and the (011) zincblende quantum
well. These systems, where the direction of the effective spin-orbit Zeeman
field is expected to be independent on the electron momentum and,
therefore, the Dyakonov-Perel mechanism of spin relaxation must be
suppressed, are considered as promising elements for the spintronics
applications. At the same time, due to the randomness of the Rashba
contribution to the spin-orbit coupling, causing an additional spin
relaxation, in both cases the relaxation time cannot exceed considerably 10
ps thus limiting the size of the base of this transistor of the order of 10 $%
\mu $m, which weakly depends on the electron momentum, and, therefore,
restricts the possibility of their experimental realization. Recent focus in
spintronics is related to Si/Ge two-dimensional structures, where the
spin-orbit coupling is weak \cite{Jantsch02,Sherman03a,Golub04,Tahan04}.
Within these systems the randomness of the SO coupling, which is related
either to the randomness in the dopant distribution as considered here or to
random bonds on the Si/Ge interfaces \cite{Golub04} will cause spin
relaxation and limit their potential applications as well.

\textit{Acknowledgment.} E.Ya. Sherman is grateful to the DARPA SpinS
program for financial support and to J.E. Sipe for very valuable discussions.



\widetext

\newpage

\begin{figure}[h]
\includegraphics{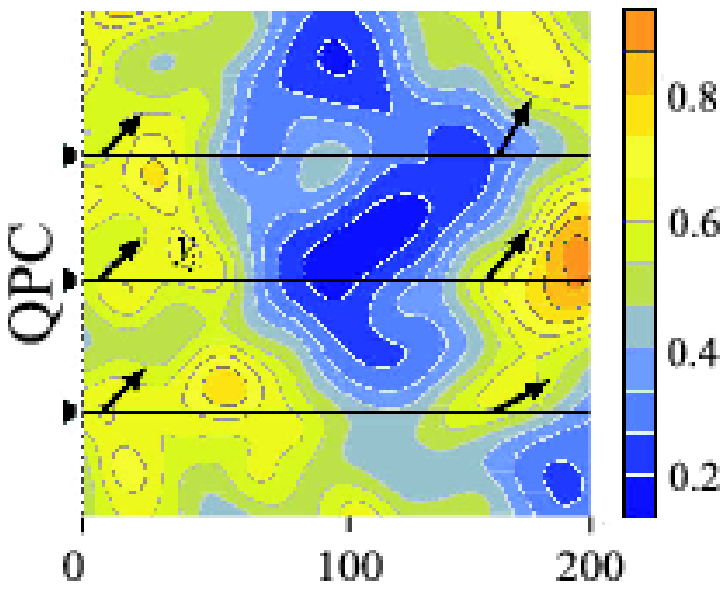}
\caption{A plot of a typical realization of the random Rashba SO coupling
(in arbitrary units) arising due to random variations in the dopant
concentration. $z_0=20$ nm, and the size of the template is $200\times200$ nm.
The quantum point contacts are marked by black triangles, thin lines show
the electron paths, and black arrows show the direction of the electron
spins.}
\label{fig:alpha2D}
\end{figure}

\newpage

\begin{figure}[h]
\includegraphics{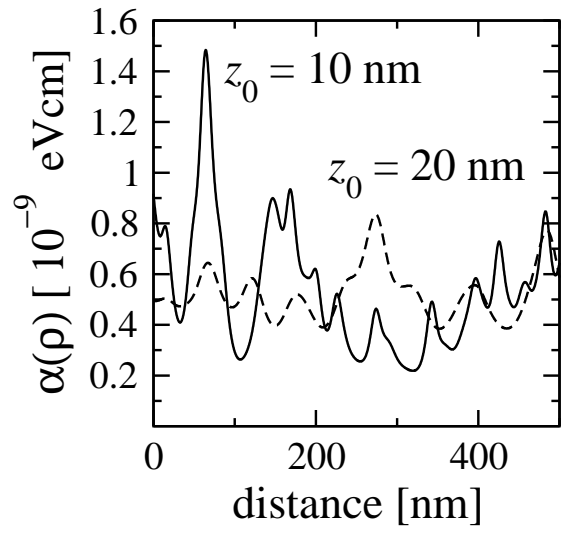}
\caption{Typical realization of the random Rashba SO coupling for different
distances between the dopant layer and the symmetry plane of the QW as
marked near the plots, $\overline{n}=2.5\times 10^{11}$ cm$^{-2}$.}
\label{fig:alphas1D}
\end{figure}

\newpage

\begin{figure}[h]
\begin{center}
\includegraphics{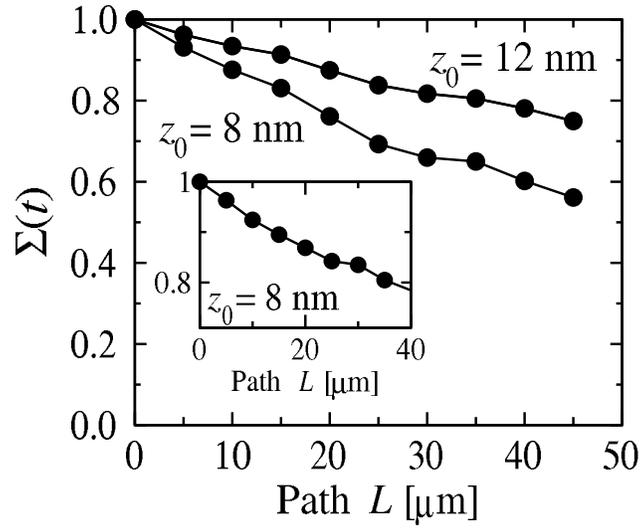}
\end{center}
\caption{Dependence of $\langle\sigma_x+\sigma_y\rangle/\protect\sqrt{2}$
on the electron path $L$, $\langle\alpha_R(V=0)\rangle=6.0\times10^{-10}$ eVcm.
The simulation includes $N=64$ electrons. It is assumed
that the regular part of the Rashba term is reduced to 0.6 of its initial
value ($\langle\alpha_R(V)\rangle=0.6\langle\alpha_R(V=0)\rangle$) by the
applied bias. The wavevector of electron $k=0.5\times 10^{6}$ cm$^{-1}$, and the
concentration of dopant ions is $\overline{n}=5\times 10^{11}$ cm$^{-2}$. The inset shows spin relaxation
at $\overline{n}=2.5\times 10^{11}$ cm$^{-2}$, $z_0=8$ nm.}
\label{fig:invariants}
\end{figure}

\newpage

\begin{figure}[h]
\begin{center}
\includegraphics{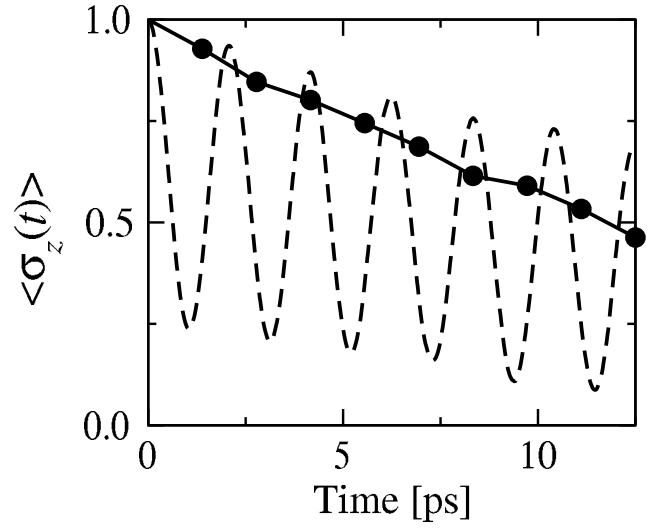}
\end{center}
\caption{Time dependence of $\langle\sigma_z\rangle$ for the (011) quantum well.
The distance between the QW symmetry plane and dopant layer is 10 nm.
The Dresselhaus SO coupling $\alpha_D=2.0\times10^{-10}$ eVcm corresponds to an 
In$_{0.5}$Ga$_{0.5}$As with $\alpha_c=25$ eV\AA$^3$ QW with $w=80$ \AA. 
The wavevector of electron $k_y=2\times 10^{6}$ cm$^{-1}$, 
$k_x=0$, $\overline{n}=2.5\times 10^{11}$ cm$^{-2}$, and $\langle\alpha_R(V=0)%
\rangle=3\times 10^{-10}$ eVcm. 
Dashed line represents $\langle\alpha_R(V)\rangle=\langle\alpha_R(V=0)\rangle/2$,
solid line represents  $\langle\alpha_R(V)\rangle=0$ (full compensation).}
\label{fig:plane}
\end{figure}

\end{document}